\documentclass[aps,prl,reprint,twocolumn,showpacs,showkeys,groupedaddress,superscriptaddress,amsmath,amssymb]{revtex4-1}
\usepackage{graphicx}
\usepackage{epic}
\usepackage{natbib}
\usepackage{color}

\begin{document}

\title{Pseudo energy wells in active systems}

\author{R. Sheshka}
\affiliation{
LITEN,  CEA-Grenoble,
17 rue des Martyrs, 38054 Grenoble,  France
}
\author{P. Recho}
\affiliation{Physico-Chimie Curie UMR 168, Institut Curie, 5 Rue Pierre et Marie Curie, 75005 Paris, France}
\affiliation{Mathematical Institute, University of Oxford, Oxford OX26GG, United Kingdom}
\author{L. Truskinovsky}
\email{trusk@lms.polytechnique.fr}
\affiliation{
LMS,  CNRS-UMR  7649,
\'{E}cole Polytechnique, 91128 Palaiseau,  France
}

\date{\today}

\begin{abstract}	
Active  stabilization in systems with zero or negative stiffness  is  an essential element of a wide variety of biological processes. We study a prototypical  example of this phenomenon at a micro-scale and show how  active rigidity, interpreted as a formation of a pseudo-well in the effective energy landscape, can be generated in an overdamped ratchet-type stochastic system.  We link the transition from negative to positive rigidity  with correlations in the noise  and show  that subtle differences in out-of-equilibrium driving may compromise the emergence of a pseudo-well. 
\end{abstract}



\maketitle

The response of a living system to mechanical loading depends not only on the  properties of the  constituents and  their connectivity,  but  also on the presence of non-thermal endogenous driving \cite{broedersz_modeling_2014}. Thus,  molecular  motors  can either stiffen  the cytoskeleton through  actively generated pre-stress \cite{koenderink_active_2009} or  fluidize it by facilitating  remodeling \cite{ranft_fluidization_2010}.  Powered by ATP hydrolysis,  living systems  can   also operate in mechanical regimes with negative  passive  stiffness as in the case of  hair cells \cite{martin_negative_2000, batters_model_2004} and   muscle half-sarcomeres \cite{vilfan_instabilities_2003, caruel_muscle_2013}. In those cases metabolic resources  are used to modify the mechanical susceptibility of the system and  stabilize the apparently unstable states \cite{schillers_real-time_2010, hawkins_stress_2014,etienne_cells_2015}. 
 
At the structural level, \emph{active rigidity} may be the outcome of tensegrity tightening \cite{ingber_tensegrity_2014}, connectivity change \cite{onck_alternative_2005}, steric interactions \cite{fletcher_cell_2010},  or the prestress exploiting strong nonlinearity of the passive response \cite{pritchard_mechanics_2014, ronceray_fiber_2015}. ATP induced stiffening can even take place at the level of individual structural elements as in the case of  the Frank-Starling effect in cardiac muscles that cannot be explained by a simple filament overlap change \cite{kobirumaki-shimozawa_cardiac_2014}. 

In this Letter we show that active rigidity  can also emerge at the micro-scale level through  resonant non-thermal excitation of molecular degrees of freedom as in the case of an inverted pendulum  \cite{butikov_improved_2011}.   Following this inertial prototype,  we construct an example of a mechanically unstable  overdamped  system where stabilization and  creation of a new  pseudo-well in the effective energy landscape can be induced by a colored noise.  The proposed mechanism of rigidity generation requires a finite distance from equilibrium and is therefore different from the more conventional entropic stabilization  \cite{vocadlo_possible_2003}.  The possibility of actively tunable rigidity   opens interesting prospects  not only in biomechanics  \cite{puglisi_cohesion-decohesion_2013}  but also in  engineering design incorporating negative  stiffness \cite{fritzen_material_2014} or  aiming at synthetic materials  stabilized dynamically \cite{bukov_universal_2015, sarkar_profiling_2015}.

We illustrate our idea on a simple bi-stable mechanical system described by a single collective variable:  the negative stiffness is viewed as  a result of  coarse-graining in a microscopic system with domineering long range interactions  \cite{caruel_mechanics_2015}. We assume that this  'snap-spring'   is exposed to  both  thermal   and   correlated  noises  and acts against a linear spring  which qualifies it as a molecular motor operating in stall conditions \cite{reimann_brownian_2002}. Instead of the conventional focus on active force, we study in this Letter  a possibility of generating by this motor active susceptibility. 

The advantage of our analytically transparent  setting  is that we can distinguish the separate effects  of  thermal (scaled with temperature $D$)  and non-thermal (scaled with affinity $A$) components of the noise on the effective  energy landscape.  We construct a non-equilibrium phase diagram in the space of parameters   $(D,A)$ showing that   bifurcations connecting different  \emph{dynamic phases} may be either sub-critical, indicating a first order phase transition, or super-critical, indicating a second order phase transition,   with the overall behavior controlled by a tri-critical point.  Some features of the  observed dynamic transitions  are  reminiscent of the behavior of the  Ising model in periodic magnetic field \cite{gallardo_analytical_2012}  and  the behavior  of folded proteins subjected to periodic forces \cite{fogle_protein_2015}.   We show however, that our system is highly sensitive to the stochastic nature of  the nonequilibrium reservoir.  Thus,  in case of  a periodic or dichotomous (DC) rocking force, a pseudo energy well exists in an extended domain of the parameter space,  while  it completely disappears if the noise is of Ornshtein-Uhlenbeck (OU) type.

\begin{figure}[h!]
\includegraphics[scale=0.9]{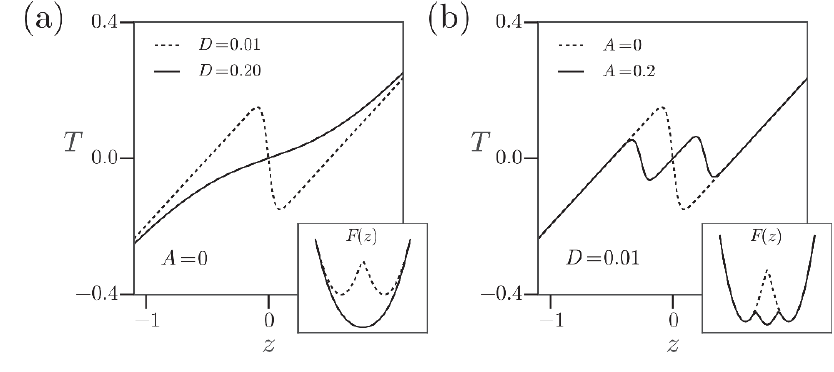}  
\caption{Force-elongation curves in the case of periodic driving (adiabatic limit).  The equilibrium system ($A=0$) is shown in  (a)  and   and out-of-equilibrium system ($A\neq 0$) - in  (b). Here $k=0.6$.}\label{Figure_1}
\end{figure}

\emph{Model.} Consider the  non-dimensional Langevin equation
 $
\dot{x} = -\partial_{x} E + \sqrt{2 D}\xi(t),
 $
where  $\xi(t)$  is a standard delta correlated noise with zero average, $D$ is a measure of temperature  and 
 $
E(x,z,t) = V(x)+k(x-z)^2/2-xf(t)
 $
is a time dependent potential. We assume that    
$V(x)=\left(\vert x \vert- 1/2\right)^2/2$ is a bi-stable potential describing the conformational change
and  $z$ is a control parameter coupled through a spring of stiffness $k$ with the internal variable $x$. The energy is  supplied to the system by  a rocking force 
$f(t)$ characterized by an amplitude $A$ and a time scale $\tau$.  The presence of a thermal noise in this mean field description suggests that the system has a finite size  \cite{paniconi_phenomenological_1997}. The external force required to maintain the system in the steady regime is
$
T (z)= k[z-\overline{\langle x\rangle}], 
$
where $\overline{\langle x\rangle}=\underset{t\rightarrow\infty}{\lim}(1/t)\int_{0}^t \int_{-\infty}^{\infty} x p(x,t) dx dt $
is the  double averaging over ensemble and  time and $p(x,t)$ is  the corresponding probability distribution.  

The main object of our study is  the  \emph{effective potential} $F(z)=\int^z T(s)ds$. While  $z$  has been introduced as a constant parameter, it can be also  viewed as a mesoscopic variable satisfying  
$
\nu \dot{z} = -\partial_{z} E+f_{ext}+ \sqrt{2 \nu  D}\xi(t)
 $
where the frictional timescale $ \nu/k \gg \tau $ and $f_{ext}$ is a slowly varying external force. As we show in  \cite{sup_2014}   the  dynamics of the ensemble and time averaged  $z$ that we denote by $Z$  is governed by 
$\nu  \dot{Z}  =-\partial_{ Z } F+\overline{ f_{ext}}$.  In the case of skeletal muscles, if $x$ characterizes the state  of a generic cross bridge,  $Z$ would  be a measure of strain at the level of the whole half-sarcomere  \cite{sup_2014}. Even in the absence of an explicit  mesoscopic variables,  the concept of an effective potential is  useful for the study of the  slow component of the motion  \cite{zaikin_doubly_2000,*baltanas_experimental_2003,*landa_nonlinear_2013,
*sarkar_controlling_2014}.

\emph{Periodic driving.} Suppose first that the driving is  square shaped   $f(t)=A(-1)^{n(t)}$ with  $n(t)=\lfloor  2t/\tau \rfloor,$ where brackets denote the integer part. An analytically transparent case is when the correlation time $\tau$ is much larger than the escape time for the bi-stable potential $V$, e.g. \cite{magnasco_forced_1993}.  For $f(t)\equiv A$  we can find the stationary solution of the Fokker-Planck equation  (with $z=const$)
$\partial_{t}p= \partial_{x}\left[
 p \partial_{x}E+D \partial_{x}p\right] 
$ analytically, compute  $\langle x(A)\rangle$ explicitly and then average the result over the period  $\langle\langle x\rangle\rangle=[\langle x(A)\rangle+\langle x(-A)\rangle]/2$, see  \cite{sup_2014} for details. The  typical force-elongation curves $T(z)$ and the corresponding potentials $F(z)$,  obtained in such adiabatic limit,  are shown in  Fig. \ref{Figure_1}.  The equilibrium system with $A=0$ exhibits negative stiffness at $z=0$ where the  effective potential $F(z)$ has a maximum (spinodal state).   
 As temperature increases  at   $A=0$  we  observe a standard entropic stabilization of the configuration $z=0$, see Fig.~\ref{Figure_1}(a),  which takes place as a second order phase transition at the equilibrium temperature $D_e=r/[8(1+k)]$  where $r$ is  a root of a transcendental equation $1+\sqrt{r/\pi }\text{e}^{-1/r}/[1+\text{erf}(1/\sqrt{r})]=r/(2k)$ \cite{sup_2014}. As the degree of non-equilibrium, characterized by $A$,  increases,  the   effective potential develops a pseudo-well with a minimum at $z=0$, see Fig.~\ref{Figure_1}(b), and  we associate this phenomenon with the emergence of active rigidity. 
  
\begin{figure}
\includegraphics[scale=1.05]{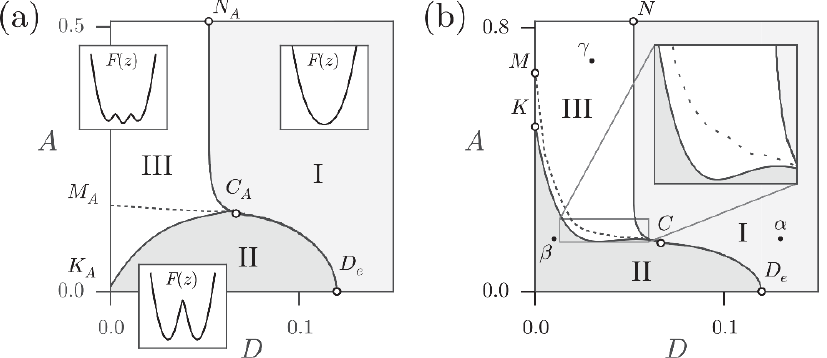}  
\caption{Phase diagram in ($A,D$) plane showing phases I,II and III: (a) - adiabatic limit, (b) - numerical solution at $\tau=100$ (b). $C_A$ is the tri-critical point, $D_e$ is the point of second order phase transion in a passive system. The Maxwell line for  first order phase transition is shown by dots. Here $k=0.6$. }\label{Figure_2}
\end{figure}
 
The  non-equilibrium steady state  (dynamic) phase diagram summarizing the  results obtained in adiabatic approximation is shown in Fig.~\ref{Figure_2} (a).  There,   the 'paramagnetic' phase I  describes the regimes where the effective potential $F(z)$ is convex,  the 'ferromagnetic' phase II  is a bi-stability domain where the potential $F(z)$ has a double well structure and,  finally, phase III is where the  function $F(z)$ has three \emph{convex} branches separated by two concave  (spinodal) regions.  If we interpret the boundary $C_A-D_e$ separating phases I and II as a line of (zero force) second order phase transitions and  the dashed line $C_A-M_A$ as a Maxwell line for the (zero force)  first order phase transition, see \cite{sup_2014},  then $C_A$ will be a  tri-critical point.  Near this point   the system can be described by the non-equilibrium Landau potential $F(z)=F_0+rz^2+qz^4+pz^6$
where the coefficients $r, q$ are the measures of passive and active excitations, respectively, while $p>0$ is a fixed parameter. Similar tri-critical point has been observed in the periodically driven mean field Suzuki-Kubo model of magnetism  \cite{tome_dynamic_1990} which can be interpreted in our terms as a  description of the $T=0$ behavior only. 

The adiabatic approximation fails at low temperatures (small $D$) where the escape time diverges and in this domain the corrected phase diagram was obtained  numerically by computing the appropriate periodic solutions of  the Fokker-Plank equation, see  Fig.~\ref{Figure_2} (b).  The high temperature part of the diagram (tri-critical point, point $D_e$ and the vertical asymptote of the boundary separating phases I and III at large values of $A$ are captured adequately by the adiabatic approximation. The new feature  is a dip of the boundary separating  Phases II and III at some $D<D_e$ leading to an interesting re-entrant behavior (cf.  \cite{van_noise_1994,pilkiewicz_reentrance_2014})  which is an effect of stochastic resonance.  To verify our numerics in the low temperature domain we  used the Kramers approximation, to show that indeed $ A=1/2$ at point $K$ and $A=1/2+k/4$ at point $M$, see \cite{sup_2014}. 

\begin{figure}[h!]
\includegraphics[scale=0.9]{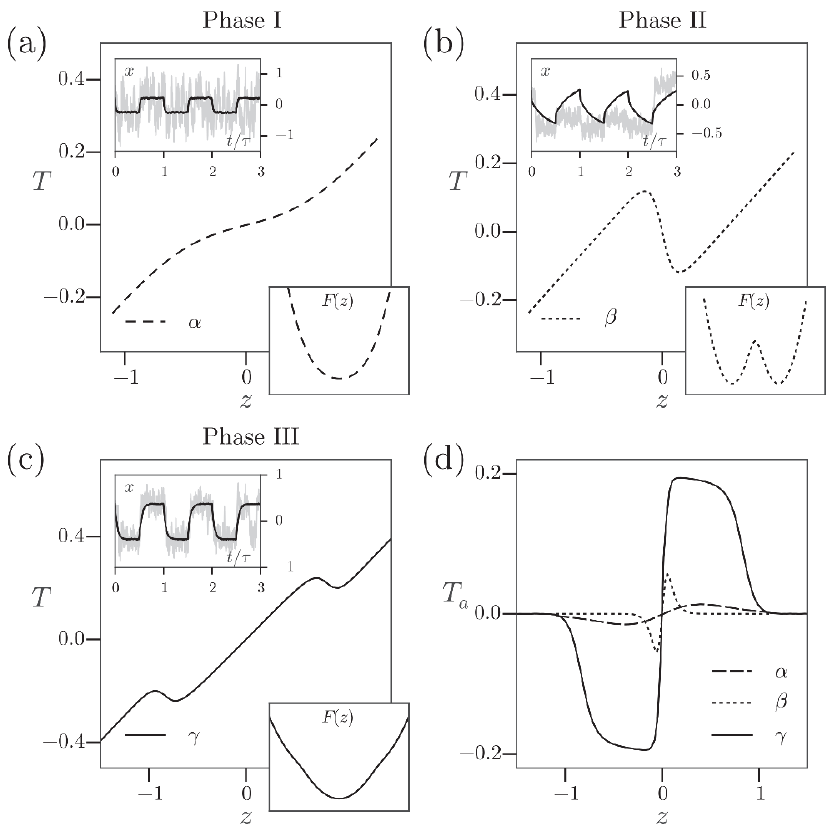}  
\caption{(a-c) Typical tension-length relations in phases I, II and III. Points $\alpha$, $\beta$ and  $\gamma$ are the same as in  Fig.~\ref{Figure_2} (b); (d)  shows the active component of the force.  Inserts illustrate the behavior of stochastic trajectories in each of the phases (gray lines)  and their  ensemble  averages (black lines) at $z\simeq0$. The other parameters: $k=0.6$, $\tau=100$. }\label{Figure_3}
\end{figure}

Force-elongation relations in different points of the ($A,D$) phase diagram  (Fig.~\ref{Figure_2} (b))  are shown in Fig.~\ref{Figure_3} where the insets illustrate  the typical  stochastic trajectories.  We observe that while in phase I thermal fluctuations  dominate  periodic driving and undermine the two wells structure of the potential,  in  phase III  the jumps between the two energy wells are fully synchronized with the rocking force. In phase II  the system  shows intermediate behavior with uncorrelated jumps between the wells.  We conclude that the  pseudo-well in phase III  has a resonant nature and remark that somewhat similar phenomena were also observed  in other driven out-of-equilibrium systems  \cite{cugliandolo_second_2000, *munoz_generic_2005, *berthier_non-equilibrium_2013}.

In Fig.~\ref{Figure_3}(d) we show the active component of the force $T_a (z)=T(z;A) -T (z;0)$  representative of  phases I, II and III.  The active contribution is significant only in phase III and  the corresponding plateau can be viewed as another signature of the presence of a pseudo-well.  Interestingly,  our prototypical device  generates  active  tension of both signs which can be interpreted as pulling at $z>0$ and pushing at $z<0$. However, in the puling regime the linear spring is stretched while in the pushing regime it is compressed. Since in biological conditions the  filaments responsible for passive stiffness would buckle in compression, e.g.  \cite{lenz_contractile_2012},  the pushing part of the active force-length relation is hardly realistic. On the other hand, the pulling part shows a striking resemblance to the isometric tetanus in skeletal muscles \cite{gordon_variation_1966}   that can be also driven  through the bi-stable potential \cite{sheshka_power-stroke-driven_2014}.

 \begin{figure}[!h]
\begin{center}
\includegraphics[scale=1.]{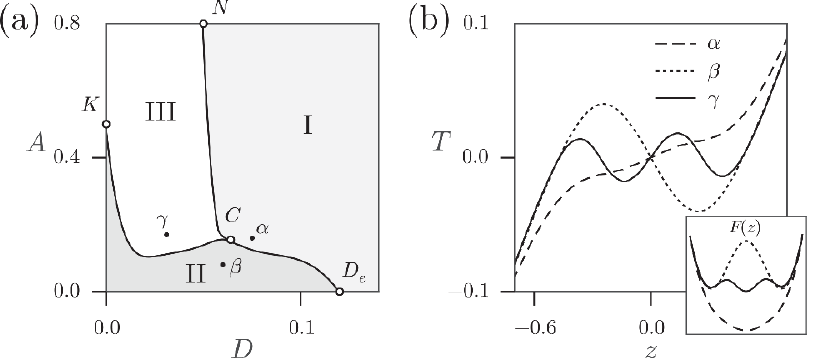}
\caption{(a) Phase diagram in the case of DC driving. The identification of phases I, II,III is the same as in Fig.~\ref{Figure_2} (a,b). (b) Typical tension-length relations in different phases (b). Here $\tau=100$ and $k=0.6$.}
\label{Figure_4}
\end{center}
\end{figure}
In view of this analogy, detailed  in \cite{sup_2014}, it is instructive to  estimate the four non-dimensional parameters  of the model  by using the available data on molecular motors operating in muscle cells. We choose the  time scale  to be $\tau^{\star}=\eta/k_{0}\sim 0.1 \text{ ms}$  where $\eta\sim  0.38\; \text{ms}.\; \text{pN}/\text{nm}$ is the viscosity adopted in \cite{caruel_muscle_2013} and $k_{0}\sim 3\; \text{pN}/\text{nm}$ is the stiffness of the cross-bridge in  pre and post power stroke configurations. The spatial scale is $l^{\star}=a\sim 10\text{nm}$, the characteristic size of a motor power-stroke \cite{linari_kinetic_2010} and the stress scale is $k^{\star}=k_{0}$. This leads to an energy scale $E^{\star}=k_{0}a^2\sim 200 \;\text{pN}.\text{nm}$. Then, the non-dimensional parameters can be estimated as follows. Parameter $k=k_{m}/k^{\star} \sim 0.6$, where  $k_{m}$ is the stiffness of the elastic part of the myosin motor \cite{lewalle_single-molecule_2008,barclay_inferring_2010}. Temperature is $D=k_{B}\Theta/E^{\star} \sim 0.01$ where  $k_{B}=4.10 \text{ pN}.\text{nm}$ is the Boltzmann constant and $\Theta \sim 300\text{K}$ the ambient temperature. For the active driving time scale, we estimate $\tau=\tau_a/\tau^{\star}\sim 100$ where  $\tau_a=40 \text{ ms}$ is the characteristic time of ATP hydrolysis \cite{howard_mechanics_2001}. Finally we take $A = \sqrt{\Delta\mu/E^{\star}} \approx 0.5$ where $\Delta\mu = 20 k_{B}\Theta$ is the degree of non-equilibrium of the hydrolysis reaction \cite{howard_mechanics_2001}. The obtained estimate ($A=0.5, D=0.01$)  suggests that  muscle myosins, operating in stall conditions (isometric contractions), are in phase III.  The proposed representation of the ATP hydrolysis (through  parameter $A$) explains stabilization of the power stroke mechanism in skeletal muscles in the negative stiffness regime \cite{caruel_muscle_2013} and   may be also behind titin based force generating mechanism at long sarcomere lengths that does not rely on actin-myosin based cross-bridge interactions \cite{schappacher_novel_2015}.
      
\emph{Dichotomous driving.} To ascertain  the robustness of these results  we now consider a different representation of the external forcing 
as a   dichotomous (DC)  or telegraphic  noise, e.g.  \cite{ichiki_singular_2012,nagai_collective_2015}. In this case $f(t)=A(-1)^{n(t)},$ where $n(t)$ is a Poisson process with ($P(n)=e^{-\lambda} \lambda^n/n!$ and   rate  parameter $\lambda = 1/(2\tau)$; we thus have 
 $\left\langle f(t)\right\rangle=A\exp(-t/\tau)$ 
and 
$\left\langle f(t),f(s)\right\rangle=A^2\exp(-|t-s|/\tau).$ The DC driven system is controlled by the same number of parameters as the periodically driven system, however, the problem  is no longer analytically tractable. The numerical solution of the ensuing stochastic differential equation  shows that the qualitative structure of the phase diagram in the ($A,D$) plane remains the same as in the case of periodic driving, see  Fig.\ref{Figure_4}. We checked our numerical results by considering an analytically tractable double limit when $\tau\rightarrow 0$,  $A\rightarrow \infty$,  while    $\tilde{D}=A^2\tau$ remains finite. In this limit phase III disappears because the system can be viewed as exposed to a white noise with effective temperature $D^*=\sqrt{D^2+\tilde{D}^2}$. Then  there is  only a second  order phase transition at the expected value of the parameter $D^*_e=r/[8(1+k)]$.  This simple limit  highlights the crucial role  of  correlations in the noise ($\tau\neq 0$).  Our next example, however, shows that correlations per se are not enough.   

\begin{figure}[!h]
\begin{center}
\includegraphics[scale=1.]{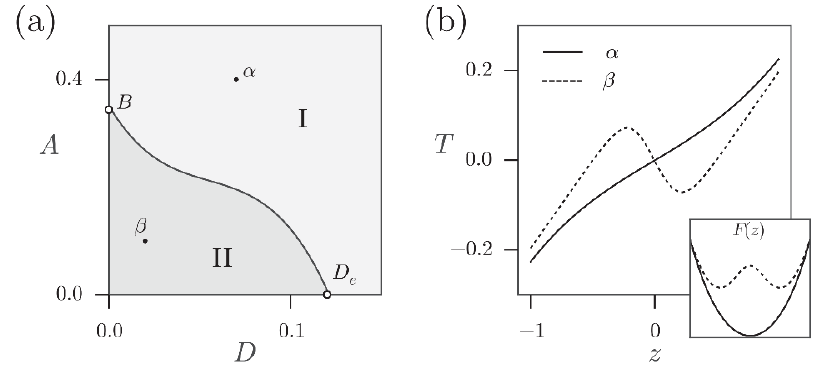}
\caption{(a) Phase diagram in the case of OU driving. The identification of phases I, II is the same as in Fig.~\ref{Figure_2} (a,b). (b) The typical tension-length relations in different phases. Here   $\tau=100$ and  $k=0.6$.}
\label{Figure_5}
\end{center}
\end{figure}

\emph{Ornstein-Uhlenbeck driving.} Suppose now that  $f(t)$ is a solution of a linear  stochastic differential equation 
 $\dot{f}= - f(t)/\tau + A\sqrt{ 2/\tau}\xi_{f}(t),$ 
where $\xi_{f}(t)$ is a standard white noise independent of $\xi(t)$. Such non-equilibrium driving is known as  Ornstein-Uhlenbeck (OU) noise, e.g.  \cite{bartussek_ratchets_1997,nagai_collective_2015} ,  and  its first ($\left\langle f(t)\right\rangle$) and second ($\left\langle f(s)f(t)\right\rangle$) moments are the same as in the case of DC if we assume, without loss of generality, that $f(0)=A$.   The Fokker-Planck equation for the probability density $p(x,f,t)$ takes the form  $ \partial_t p = \partial_x(p\partial_x E+D\partial_{x} p)+\tau^{-1}\partial_f( f  p+A^{2} \partial_{f}p).$ By solving it numerically we obtain a phase diagram shown in Fig.~\ref{Figure_5}(a).  A striking  feature of this diagram is that phase III is  missing  because, in contrast to periodic and DC case,  the noise is now unbounded and the system can always escape from a neighborhood of a resonant state. The behavior  of the force-elongation relations  shown in Fig.~\ref{Figure_5}(b) is compatible with the idea of purely entropic stabilization, in particular,  the  limit of thermal noise  is again recovered when  $\tau\rightarrow 0$ and  $A\rightarrow \infty$, with $\tilde{D}=A^2\tau$ fixed. 

\begin{figure}
\includegraphics[scale=1]{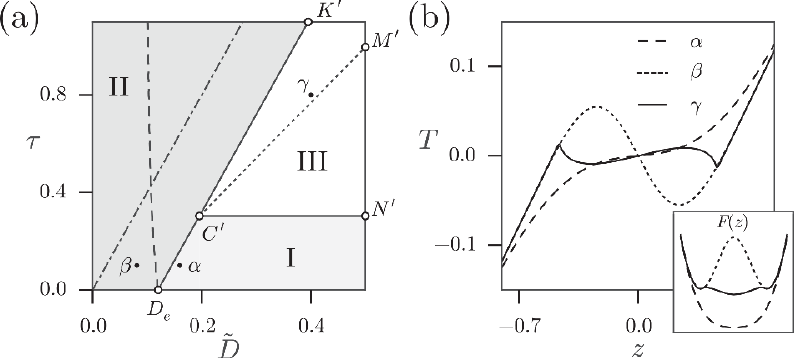}  
\caption{(a) Phase diagram in the case of DC driving. The identification of phases I, II and III is the same as in Fig.~\ref{Figure_2} (a,b). The dash-dotted line  - the boundary between  phases II and III (periodic driving); the dashed line  - the boundary between Phases I and II (OU driving). (b) Typical tension-length relations in the case of DC driving in different phases (b). Parameters $k=0.6$, $D=0$.}\label{Figure_6}
\end{figure}

\emph{Zero temperature limit.}
To clarify further  the differences between our three representations of a non-equilibrium bath,  we  compare in all three cases the ($\tau, \tilde{D}$) phase diagrams corresponding to the limit  $D\rightarrow 0$  where the thermal component of the noise is absent.  
In the DC case the solution of the limiting Fokker-Plank equation can be written explicitly \cite{hanggi_colored_1995} 
$$
p_{DC}(x)\sim Q(x)^{-1} \exp\left(-\int^x\frac{\partial_y\tilde{V}(y)/\tau}{A^2-(\partial_y\tilde{V}(y))^2}dy\right) .
$$
where $Q =A^2-(\partial_x\tilde{V}(x))^2$ and $\tilde{V} =V+k(x-z)^2/2$. The choice  of the normalization constant depends on the parameters and is detailed in \cite{sup_2014}. The resulting phase diagram, shown in Fig.~\ref{Figure_6}(a), exhibits  all three phases  with a  tri-critical point  $C'$  located at 
$\tau_{C'} = [2(k+1)]^{-1}$ and $\tilde{D}_{C'} =D_e+[2(k+1)]^{-1}/4$. The behavior of the force-elongation relations in different phases  is illustrated in Fig.~\ref{Figure_6}(b). 

In the case of OU driving with $D=0$ an analytical approximation of the stationary probability distribution is available for small $\tau$ only \cite{hanggi_colored_1995}
$$
p_{OU}(x)\sim R(x)\exp\left(-\frac{\tilde{V}(x)+\tau (\partial_x\tilde{V}(x))^2/2}{A^2\tau}\right)
$$
where $R(x)=|1+\tau \partial_{xx}\tilde{V}(x)|$, see  \cite{sup_2014} for the details. The resulting phase diagram does not contain phase III and the line dividing phases I and II is shown in Fig.~\ref{Figure_6}(a)  (dashed line). 
The problem with periodic driving exhibits  in the limit $D\rightarrow 0$  only phases II and III  even for rapidly oscillating external fields, see  the dash-dotted line in Fig.~\ref{Figure_6}(a). In this perspective,  the DC driving emerges as an intricate amalgam of OU and periodic noises with none of them dominating the other. 

\emph{Conclusions.} To complement the existing microscopic models of  force  generation (Brownian ratchets),  we proposed a  conceptually similar  model of  rigidity generation (Brownian snap-springs). The model,  invoking some interesting parallels between condensed matter physics and biomechanics,  shows that by controlling the degree of non-equilibrium in the system, one can modify the structure of the effective energy landscape.  In particular, this implies that unstable or marginally stable mechanical states may be  stabilized  by out-of-equilibrium  ATP hydrolysis reaction.  Our results also suggest that the mechanical action of a non-equilibrium reservoir can  be crucially sensitive to the higher moments of the stochastic noise.   

 The authors thank J.-F. Joanny, R. Garc\'{i}a Garc\'{i}a and M. Caruel for helpful discussions. 

\bigskip 
\end{document}